\documentclass[twocolumn,aps,pra,amsmath,amssymb]{revtex4-1}
\usepackage[latin9]{inputenc}
\usepackage{mathrsfs}
\usepackage{amstext}
\usepackage{amssymb}
\usepackage{graphicx}
\usepackage{esint}

\makeatletter
\usepackage{epsfig}
\usepackage{bm}
\usepackage{dcolumn}
\usepackage{color}

\makeatother

\begin{document}
\title{Polariton-polariton interaction beyond the Born approximation: A toy
model study}
\author{Hui Hu$^{1}$, Hui Deng$^{2,3}$, and Xia-Ji Liu$^{1}$}
\affiliation{$^{1}$Centre for Quantum Technology Theory, Swinburne University
of Technology, Melbourne, Victoria 3122, Australia}
\affiliation{$^{2}$Department of Physics, University of Michigan, Ann Arbor, MI
48109, USA}
\affiliation{$^{3}$Applied Physics Program, University of Michigan, Ann Arbor,
MI 48109, USA}
\date{\today}
\begin{abstract}
We theoretically investigate the polariton-polariton interaction in
microcavities beyond the commonly used Born approximation (i.e., mean-field),
by adopting a toy model with a contact interaction to approximately
describe the attraction between electrons and holes in quantum well
and by using a Gaussian pair fluctuation theory beyond mean-field.
We obtain a density or chemical potential independent polariton-polariton
interaction strength even in two-dimensions, which result from coupling
to the photon field. We show that quantum fluctuations lead to about
a factor of two reduction in the polariton-polariton interaction strength
within our toy model. Together with corrections to the 1s exciton
approximation at very strong light-matter coupling, we find the polariton-polariton
interaction strength under typical experimental conditions is overestimated
by a factor three in the widely used theories, if our toy model can
qualitatively simulate the polariton interaction in GaAs quantum wells.
We compare our prediction with the most recent measurement and argue
that the beyond-Born-approximation effect to the polariton-polariton
interaction strength is crucial for a quantitative understanding of
the experimental data by E. Estrecho \textit{et al.}, Phys. Rev. B
\textbf{100}, 035306 (2019). 
\end{abstract}
\maketitle

\section{Introduction}

Exciton-polaritons in microcavities are half-light and half-matter
bosonic quasi-particles, arising from the strong coupling between
the photo field and tightly-bound electron-hole pairs (i.e., excitons)
\cite{Deng2010,Byrnes2014}. Due to the ultra-small effective mass
inherent from the light, Bose-Einstein condensation (BEC) of exciton-polaritons
can occur at high temperatures \cite{Deng2002,Schneider2017}. Together
with the nonlinearity originating from their underlying ferminoic
constituents, exciton-polaritons provide an attractive platform to
realize new technologies such as efficient and ultrafast optical switches
and optical transistors \cite{Fraser2016,Sanvitto2016}.

Due to the critical role of the polariton nonlinearity in phase transitions
and nonlinear optical device concepts, there have been intense experimental
\cite{Ferrier2011,Brichkin2011,Kim2016,Sun2017,MunozMatutano2019,Delteil2019,Estrecho2019,Hu2019arXiv}
and theoretical effort \cite{Ciuti1998,Tassone1999,Glazov2009,Xue2016,Levinsen2019}
to characterize the polariton nonlinearity over the past few decades.
However, there continues to be conceptual difficulties in understanding
and calculating the polariton nonlinearity with the widely used mean-field
approach. The mean-field approach produces a constant polariton-polariton
interaction strength $g_{PP}$, or, a linearly increasing interaction
energy with polariton density. This linear density dependence, however,
is not anticipated for weakly-interacting two-dimensional (2D) Bose
gases \cite{NoteShortRangeInteraction}. According to the Bogoliubov
theory, the relation between the chemical potential $\mu_{B}$ and
the number density $n$ of an interacting 2D Bose gas would be given
by \cite{Schick1971,Mora2009}, 
\begin{equation}
n\simeq\frac{m_{B}\mu_{B}}{4\pi\hbar^{2}}\ln\left[\frac{4\hbar^{2}}{m_{B}\mu_{B}a_{s}^{2}e^{2\gamma+1}}\right],\label{eq:EoS2DBoseGas}
\end{equation}
where $m_{B}$ is the mass of bosons, $a_{s}$ is the 2D $s$-wave
scattering length for the \emph{short-range} (contact) interaction
between bosons \cite{NoteShortRangeInteraction}, and $\gamma\simeq0.577$
is Euler's constant. This indicates a density or chemical potential
dependent interaction strength 
\begin{equation}
g\left(\mu_{B}\right)=\frac{\mu_{B}}{n}=\frac{4\pi\hbar^{2}}{m_{B}}\ln^{-1}\left[\frac{4\hbar^{2}}{m_{B}\mu_{B}a_{s}^{2}e^{2\gamma+1}}\right].
\end{equation}
In particular, towards the dilute limit the interaction strength would
vanish due to the vanishingly small chemical potential and density.
This result apparently disagrees with the linear dependence observed
or assumed in experiments, if we treat polaritons as a gas of weakly
interacting bosons. In greater detail, to date most calculations of
the polariton-polariton interaction strength are based on the Born
approximation \cite{NoteBornApproximation}. In the exciton-polariton
model, it leads to a polariton-polariton interaction strength \cite{Tassone1999},
\begin{equation}
g_{PP}^{(0)}=X_{LP}^{4}g_{XX}^{(0)},\label{eq:gPP0EP}
\end{equation}
where 
\begin{equation}
g_{XX}^{(0)}\simeq6.06E_{X}a_{X}^{2}=6.06\hbar^{2}/M\label{eq:gXX2D}
\end{equation}
is the constant exciton-exciton interaction strength in 2D and 
\begin{equation}
X_{LP}^{2}=\frac{1}{2}\left[1+\frac{\delta/2}{\sqrt{\delta^{2}/4+\Omega^{2}}}\right]\label{eq:XLP2}
\end{equation}
is the excitonic Hopfield coefficient with the photon detuning $\delta$
(measured with respect to the exciton energy $-E_{X}$) and with the
light-matter coupling $\Omega$. Here, $a_{X}$ and $E_{X}=\hbar^{2}/(Ma_{X}^{2})$
are the Bohr radius and binding energy of excitons, respectively,
and we have assumed for simplicity that electrons and holes take the
same mass $m_{\textrm{e}}=m_{\textrm{h}}=m_{\textrm{eh}}=M$. We have
also used the superscript ``0'' to explicitly indicate the results
within the Born approximation. Eq. (\ref{eq:gPP0EP}) is very easy
to understand since the interaction between polaritons is mediated
by the excitonic component of polaritons only. However, it should
be corrected when the light-matter coupling $\Omega$ becomes strong
and comparable to $E_{X}$, so that the standard exciton-polariton
model starts to break down. This non-trivial effect due to strong
light-matter coupling is well-known in the literature \cite{Brichkin2011,Tassone1999}
and most recently has been rigorously treated by solving the exact
two-body problem of the underlying fermionic electron-hole-photon
Hamiltonian in the dilute limit \cite{Levinsen2019}. It was shown
that the correction to Eq. (\ref{eq:gPP0EP}) can be about $\sim20\%$
at the very strong coupling regime when $\Omega\sim E_{X}$ \cite{Levinsen2019}.
On the other hand, experimentally, the exciton-exciton interaction
strength $g_{XX}^{(0)}$ in Eq. (\ref{eq:gPP0EP}) may also need revision,
considering the quasi-2D configuration of the quantum well, whose
width $l_{z}$ would be similar to $a_{X}$ \cite{Estrecho2019}.
In such a situation, a rough estimation gives rise to, 
\begin{equation}
g_{XX,\textrm{q2d}}^{(0)}=\frac{26\pi}{3}E_{X}a_{X}^{2}\left(\frac{a_{X}}{l_{z}}\right).\label{eq:gXXquasi2D}
\end{equation}
This expression was used by Estrecho and his collaborators to set
a theoretical \emph{upper} bound for the polariton-polariton interaction
strength \cite{Estrecho2019}. It is about three times larger than
the measured value.

It is certainly not satisfactory to restrict theoretical analysis
just to the Born approximation. This is particularly relevant in 2D,
where quantum and thermal fluctuations are so significant that the
equation of state of the system can qualitatively be altered \cite{He2015}.
The density or chemical potential interaction strength of an interacting
2D Bose gas mentioned in the above is already an excellent example.
Even in three dimensions (3D), the beyond-Born-approximation effect
could be very significant. A well-known case is a two-component ultracold
atomic Fermi gas with a contact interaction characterized by a 3D
$s$-wave length length $a_{F}$. In the BEC limit where tightly bound
molecules are formed, the exact molecule-molecule scattering length
is $a_{s}\simeq0.6a_{F}$ \cite{Petrov2004,Brodsky2006}, much smaller
than the result $a_{s}^{(0)}=2a_{F}$ obtained within the Born approximation.

In this work, we aim to better understand the polariton-polariton
interaction in 2D by going beyond the Born approximation. This is
possible if we replace the Coulomb interaction between electrons and
holes with a short-range contact interaction, whose scattering length
is tuned to correctly reproduce the binding energy of excitons. Therefore,
we are able to construct a \emph{toy} model for the electron-hole-photon
system, which captures the important underlying \emph{fermionic} degree
of freedom of exciton-polaritons. By applying a Gaussian pair fluctuation
theory (GPF) beyond mean-field as in the previous investigation of
ultracold atoms \cite{Hu2020,NotationNote}, we reliably calculate
the polariton-polariton interaction strength at various light-matter
couplings and photon detunings for the toy model.

Two main observations are worth noting. Firstly, in the presence of
the photon field, the scattering of two composite bosons (i.e., excitons)
is strongly modified. In particular, at strong light-matter coupling,
where the photon field is notably populated, the internal fermionic
degree of freedom of excitons can not be ignored. The modification
to the exciton-exciton scattering due to the photon field provides
the correct theoretical understanding why a nearly constant, density
independent polariton-polariton interaction strength was found in
the experiments \cite{Estrecho2019}. Secondly, the effect beyond
the Born approximation is significant and typically leads to about
a factor of two reduction in the interaction strength. Combined with
the non-trivial effect due to strong light-matter coupling, in total
we find that the polariton-polariton interaction strength under typical
experimental conditions to be about a factor of three smaller relative
to the prediction of Eq. (\ref{eq:gPP0EP}).

We note that, for a small light-matter coupling, we may use a purely
bosonic model Hamiltonian to describe the exciton-polariton system
\cite{Hu2020arXiv}. In that case, the beyond mean-field effect can
be captured by using the Bogoliubov theory, which takes into account
the many-body effects and strong quantum fluctuations in two dimensions
\cite{Hu2020arXiv}, and momentum-dependent interactions may also
be used without the simplification to contact interactions. Our GPF
results from the fermionic toy model agree well with the analytic
Bogoliubov predictions obtained with the bosonic exciton-polariton
model, if we use the same parameters under the same condition.

The rest of the paper is organized as follows. In the next chapter
(Sec. II), we briefly review the GPF theory of the toy model with
a contact electron-hole interaction for the exciton-polariton system
in microcavities. In Sec. III, we consider the case with a small light-matter
coupling and a large photon detuning, for which a weakly interacting
2D exciton condensate is recovered. We discuss the exciton-exciton
interaction within the Born approximation (i.e., mean-field level)
and beyond the Born approximation (i.e., GPF level). In Sec. IV, we
investigate the polariton system at large light-matter couplings and
define a generalized excitonic Hopfield coefficient, which captures
the oscillator strength saturation effect and the reduced size of
exciton wave-functions due to the photon-mediated attraction \cite{Levinsen2019}.
We show that the correction to the polariton-polariton interaction
strength beyond the Born approximation might be characterized by using
the mean-field density fractions. In Sec. V, we assume the insensitivity
of the beyond-Born-approximation effect on the underlying electron-hole
attraction and compare our prediction with the latest measurement
of the polariton-polariton interaction strength \cite{Estrecho2019}.
Finally, we summarize in Sec. VI.

\section{Theoretical model and Gaussian pair fluctuation theory}

The 2D electron-hole-photon system in microcavities can be described
by the model Hamiltonian $\mathscr{H}=\mathscr{H}_{0}+\mathscr{H}_{\textrm{LM}}+\mathscr{H}_{\textrm{C}}$
as \cite{Kamide2010,Byrnes2010,Yamaguchi2012} 
\begin{eqnarray}
\mathscr{H}_{0} & = & \sum_{\mathbf{k}\sigma}\xi_{\mathbf{k}}c_{\mathbf{k}\sigma}^{\dagger}c_{\mathbf{k}\sigma}+\sum_{\mathbf{q}}\left[\frac{\hbar^{2}\mathbf{q}^{2}}{2m_{\textrm{ph}}}+\delta_{0}-\mu\right]\phi_{\mathbf{q}}^{\dagger}\phi_{\mathbf{q}},\\
\mathscr{H}_{\textrm{LM}} & = & \frac{g_{0}}{\sqrt{\mathcal{S}}}\sum_{\mathbf{kq}}\left[\phi_{\mathbf{q}}^{\dagger}c_{\frac{\mathbf{q}}{2}-\mathbf{k}h}c_{\frac{\mathbf{q}}{2}+\mathbf{k}e}+\textrm{h.c.}\right],\\
\mathscr{H}_{\textrm{C}} & = & \frac{1}{2\mathcal{S}}\sum_{\mathbf{k}\mathbf{k}'\mathbf{q}}^{\sigma\sigma'}V_{\mathbf{k}\mathbf{k}'}^{\sigma\sigma'}c_{\frac{\mathbf{q}}{2}+\mathbf{k}\sigma}^{\dagger}c_{\frac{\mathbf{q}}{2}-\mathbf{k}\sigma'}^{\dagger}c_{\frac{\mathbf{q}}{2}-\mathbf{k}'\sigma'}c_{\frac{\mathbf{q}}{2}+\mathbf{k}'\sigma}.\label{eq: HamiCoul}
\end{eqnarray}
Here, $\xi_{\mathbf{k}}\equiv\hbar^{2}\mathbf{k}^{2}/(2M)-\mu/2$,
$\delta_{0}$, $\mu$, $g_{0}$ and $\mathcal{S}$ are the electronic
dispersion within an effective mass approximation, \emph{bare} cavity
detuning, chemical potential, \emph{bare} light-matter coupling strength,
and the area of the system, respectively. We have taken the same mass
$M=m_{\textrm{eh}}\simeq0.067m_{0}$ for electrons and holes (where
$m_{0}$ is the free-electron mass) and an ultra-small photonic mass
$m_{\textrm{ph}}\simeq3\times10^{-5}m_{0}$ due to the microcavity
confinement \cite{Deng2010}. $c_{\mathbf{k}\sigma}$ are the annihilation
operators of electrons ($\sigma=e$) and holes ($\sigma=h$), and
$\phi_{\mathbf{q}}$ denote the annihilation operators of photons.

In Eq. (\ref{eq: HamiCoul}), $V_{\mathbf{k}\mathbf{k}'}^{\sigma\sigma'}$
are the Coulomb-like interactions among electrons and holes, and are
defined as the Fourier transformation of a screened potential \cite{Keldysh1979,Cudazzo2011},
\begin{equation}
V_{C}^{\sigma\sigma'}\left(r\right)=\chi_{\sigma\sigma'}\frac{\pi e^{2}}{2\varepsilon_{s}r_{0}}\left[H_{0}\left(\frac{r}{r_{0}}\right)-Y_{0}\left(\frac{r}{r_{0}}\right)\right],\label{eq:VC}
\end{equation}
where $\chi_{\sigma\sigma'}=+1$ for $\sigma=\sigma'$ and $\chi_{\sigma\sigma'}=-1$
for $\sigma\neq\sigma'$, $\varepsilon_{s}$ is the dielectric constant
of the substrate surrounding the quantum well, $H_{0}(x)$ and $Y_{0}(x)$
are respectively the Struve and Neumann functions, and $r_{0}$ is
an effective screening length. This particular form of the Coulomb-like
interaction is due to the large difference in the dielectric constants
of the quantum well and of the substrate, which strongly modifies
the Coulomb interaction at short distance \cite{Keldysh1979,Cudazzo2011}.
The model Hamiltonian is extremely difficult to solve because of the
\emph{non-local} nature of the Coulomb interaction. To find a way
around, we propose a \emph{toy} model by replacing the Coulomb interaction
with a local contact interaction \cite{Yamaguchi2012,Hanai2018},
i.e., 
\begin{equation}
\mathscr{H}_{\textrm{C}}=\frac{u_{0}}{\mathcal{S}}\sum_{\mathbf{k}\mathbf{k}'\mathbf{q}}c_{\frac{\mathbf{q}}{2}+\mathbf{k}e}^{\dagger}c_{\frac{\mathbf{q}}{2}-\mathbf{k}h}^{\dagger}c_{\frac{\mathbf{q}}{2}-\mathbf{k}'h}c_{\frac{\mathbf{q}}{2}+\mathbf{k}'e},\label{eq:HamiContact}
\end{equation}
where the interaction strength $u_{0}$ should be tuned to reproduce
the correct ground-state energy of excitons with the Coulomb-like
interaction Eq. (\ref{eq:VC}).

It is useful to note that, in ultracold atomic physics our toy model
Hamiltonian describes a two-component interacting Fermi gas near Feshbach
resonances at the crossover from a BEC to a Bardeen--Cooper--Schrieffer
(BCS) superfluid \cite{Ohashi2002,Ohashi2003,Liu2015,Hu2019}. The
Feshbach coupling is simply the light-matter coupling here. The photons
now play the role of the closed-channel molecules, if we ignore a
small modification to the photon mass (i.e., we cannot have the relation
$m_{\textrm{ph}}=2M$, which holds for ultracold atoms), while the
excitons at low density correspond to the tightly-bound Cooper pairs
in the open channel. For more details, we refer to the discussions
in Ref. \cite{Hu2020} and Ref. \cite{Yamaguchi2012}. At a broad
Feshbach resonance, which is realized when the light-matter coupling
is \emph{infinitely} strong, our toy model Hamiltonian has actually
been investigated both experimentally \cite{Makhalov2014,Ries2015,Turlapov2017}
and theoretically \cite{Levinsen2015,Mulkerin2015}. Here, the purpose
of this work is to understand the molecular scattering length in the
case of a very strong yet finite light-matter coupling or Feshbach
coupling, which is not explored so far in the context of ultracold
atoms.

The use of contact interactions both for the electrons and holes ($u_{0}$)
and for the light-matter coupling ($g_{0}$) will lead to an \emph{ultraviolet}
divergence. This divergence can be formally removed by the so-called
regularization procedure, after which the bare parameters $u_{0}$,
$g_{0}$, and $\delta_{0}$ will be replaced by $u$, $g$, and $\tilde{\delta}=\delta-E_{X}=E_{\textrm{cav}}$,
respectively. Here, $E_{\textrm{cav}}$ is the cavity energy measured
from the edge of the bandgap, and the renormalized parameters $u$
and $g$ are explicitly related to the physical observables of the
exciton binding energy $E_{X}$ and the Rabi coupling $\Omega$ as
follows \cite{Hu2020}: 
\begin{eqnarray}
u & = & \frac{4\pi\hbar^{2}}{M}\ln^{-1}\left(\frac{E_{X}}{\varepsilon_{0}}\right),\\
g & = & 2\sqrt{\pi}\Omega a_{X}\ln^{-1}\left(\frac{E_{X}}{\varepsilon_{0}}\right),
\end{eqnarray}
where $\varepsilon_{0}\ll E_{X}$ is an unimportant energy scale used
to regularize the logarithmic \emph{infrared} divergence commonly
encountered in two dimensions. For more details on the renormalization,
we refer to Supplemental Material of Ref. \cite{Hu2020}, which also
explains the solution of the two-particle problem.

To obtain the polariton-polariton interaction strength (which is intrinsically
a six-particle problem, involving two photons, two electrons and two
holes), we solve our toy model Hamiltonian using the many-body GPF
theory \cite{He2015,Hu2006,Hu2007,Diener2008} and then consider the
low-density dilute limit. The details of the GPF formalism are again
outlined in Ref. \cite{Hu2020}. Here, for self-containedness we briefly
review the main equations. Taking the Hubbard--Stratonovich transformation,
we first introduce a pairing field to decouple $\mathscr{H}_{\textrm{C}}$
in Eq. (\ref{eq:HamiContact}) and integrate out the fermionic fields
$c_{\mathbf{k}\sigma}$. We then obtain an effective action for the
pairing field and photon field, whose superposition could be understood
as a polariton field. At zero temperature, the saddle-point solution
of the polariton field gives rise to a mean-field thermodynamic potential
\cite{Hu2020}, 
\begin{equation}
\Omega_{\textrm{MF}}=-\frac{\Delta^{2}}{u_{\textrm{eff}}}+\sum_{\mathbf{k}}\left[\xi_{\mathbf{k}}-E_{\mathbf{k}}+\frac{\Delta^{2}}{\hbar^{2}\mathbf{k}^{2}/M+\varepsilon_{0}}\right],\label{eq:OmegaMF}
\end{equation}
where $\Delta$ is an order parameter satisfying the gap equation
$\partial\Omega_{\textrm{MF}}/\partial\Delta=0$, 
\begin{equation}
u_{\textrm{eff}}\equiv u+\frac{g^{2}}{\mu-\tilde{\delta}}
\end{equation}
is an effective interaction incorporating the photon-mediated attraction,
and $E_{\mathbf{k}}\equiv\sqrt{\xi_{\mathbf{k}}^{2}+\Delta^{2}}$
is the dispersion relation for fermionic Bogoliubov quasi-particles.
To go beyond mean-field, we expand the effective action around the
saddle point and keep the bilinear terms in the polariton field (i.e.,
the so-called Gaussian fluctuations) \cite{Hu2006,Hu2007,Diener2008,Keeling2005}.
Integrating out these fluctuations, we obtain the GPF thermodynamic
potential from quantum fluctuations \cite{Hu2020}, 
\begin{equation}
\Omega_{\textrm{GPF}}=\frac{1}{2}k_{B}T\sum_{\mathcal{Q}}\ln\det\Gamma\left[\mathcal{Q}=\left(\mathbf{q},i\nu_{n}\right)\right]e^{i\nu_{n}0^{+}},\label{eq:OmegaGPF}
\end{equation}
where $\Gamma(\mathcal{Q})$ with bosonic Matsubara frequencies $\nu_{n}=2\pi nk_{B}T$
($n\in\mathbb{Z}$) is the Green function of the polariton field.
It is a 2 by 2 matrix with off-diagonal terms representing the phase
correlation of the superfluid. In the normal phase above the superfluid
transition temperature, the off-diagonal terms disappear and the diagonal
term becomes a scalar variable \cite{Hu2020}, 
\begin{equation}
\Gamma\left(\mathcal{Q}\right)=\left[\frac{1}{u_{\textrm{eff}}\left(Q\right)}+\Pi\left(\mathcal{Q}\right)\right]^{-1},\label{eq:VertexFunction}
\end{equation}
where $u_{\textrm{eff}}(\mathcal{Q})\equiv u+g^{2}/[i\nu_{n}-\hbar^{2}\mathbf{q}^{2}/(2m_{\textrm{ph}})+\mu-\tilde{\delta}]$
is a momentum- and frequency-dependent effective interaction strength,
and $\Pi(Q)$ is the pair propagator. In the vacuum limit (i.e., the
two-particle limit), the pair propagator takes the form \cite{Hu2020},
\begin{equation}
\Pi_{\textrm{vac}}\left(Q\right)=-\frac{M}{4\pi\hbar^{2}}\ln\left[\frac{\hbar^{2}q^{2}/\left(4M\right)-i\nu_{n}}{\varepsilon_{0}}\right].\label{eq:PairPropagatorVacuum}
\end{equation}
By substituting the above vacuum pair propagator into Eq. (\ref{eq:VertexFunction}),
we can determine the pole of the polariton Green function and obtain
the dispersion relation of the polaritons in the dilute limit, which
consists of two branches: the lower-polariton branch $E_{LP}(\mathbf{q})$
and the upper-polariton branch $E_{UP}(\mathbf{q})$ \cite{Hu2020}.

The GPF theory of exciton-polaritons is easy to numerically implement.
For a given chemical potential, we determine the order parameter using
the gap equation. The mean-field and GPF thermodynamic potentials
are then calculated, from which we obtain the total carrier densities
$n_{\textrm{tot}}=n_{\textrm{MF}}+n_{\textrm{GPF}}$, where 
\begin{eqnarray}
n_{\textrm{MF}} & = & -\frac{\partial\Omega_{\textrm{MF}}}{\partial\mu},\\
n_{\textrm{GPF}} & = & -\frac{\partial\Omega_{\textrm{GPF}}}{\partial\mu}.
\end{eqnarray}

One advantage of our GPF theory is that it can provide a reliable
equation of state at zero temperature \cite{He2015,Hu2006,Diener2008}.
In particular, in the dilute limit, where the chemical potential depends
linearly on the density (i.e., the linear regime), it gives an approximate
but reasonably accurate molecular scattering length. For example,
for a two-component interacting Fermi gas at BEC-BCS crossover in
three dimensions, the molecular scattering length predicted by the
GPF theory is about $a_{s}\simeq0.55a_{F}$ \cite{Hu2006,Diener2008},
which is slightly smaller than the exact value $a_{s}\simeq0.60a_{F}$
\cite{Petrov2004}. In two dimensions of interest, the GPF theory
also provides a very accurate molecular scattering length \cite{He2015},
as we shall discuss in detail in the next section.

\section{2D exciton condensate with contact interactions}

For an interacting 2D Fermi gas with a contact interaction in the
BEC limit, the system can be viewed as a weakly interacting Bose gas
of molecules \cite{He2015,Makhalov2014}, with mass $m_{B}=2M$ and
density $n=n_{F}/2$ ($n_{F}$ is the density of fermions). The exact
four-body calculation shows that the molecular scattering length $a_{s}$
is related to the 2D scattering length between fermions $a_{2D}$
through \cite{Petrov2003} 
\begin{equation}
a_{s}=\kappa a_{2D}\simeq0.56a_{2D}.
\end{equation}
Here, $a_{2D}=2e^{-\gamma}a_{X}$ can be calculated by using the binding
energy $E_{X}=4\hbar^{2}/(Ma_{2D}^{2}e^{2\gamma})$. The molecular
scattering length determined from the 2D GPF theory \emph{coincides}
with the exact value if we keep the two significant digits \cite{He2015}.
According to the Bogoliubov theory of a 2D weakly interacting Bose
gas, Eq. (\ref{eq:EoS2DBoseGas}), we thus obtain, 
\begin{equation}
n\simeq\frac{1}{2\pi a_{X}^{2}}\left[-2\ln\kappa-\left(\ln2+1\right)-\ln\frac{\mu_{B}}{E_{X}}\right]\frac{\mu_{B}}{E_{X}}.\label{eq:excitonGPF}
\end{equation}
In contrast, the mean-field theory cannot predict qualitatively correct
equation of state. By writing the molecular chemical potential $\mu_{B}$
in terms of the chemical potential of fermions $\mu_{F}=\mu/2$ (i.e.,
$\mu_{B}=2\mu_{F}+E_{X}=\mu+E_{X}$), from the mean-field equation
of state \cite{He2015}
\begin{equation}
\mu_{F}+\frac{E_{X}}{2}=\varepsilon_{F}\equiv\frac{\hbar^{2}\left(2\pi n_{F}\right)}{2M},
\end{equation}
we find that,

\begin{equation}
\mu_{B}=\frac{4\pi\hbar^{2}n}{M}=4\pi E_{X}a_{X}^{2}n,\label{eq:excitonMF}
\end{equation}
implying a molecule-molecule interaction strength $g_{m}=4\pi E_{X}a_{X}^{2}$
within the Born approximation.

In the case that the photon field is not occupied, our toy model describes
exactly the 2D interacting Fermi gas and molecules discussed in the
above can be viewed as excitons. Hence, we find that the exciton-exciton
interaction strength in the toy model within the Born approximation
is, 
\begin{equation}
g_{XX}^{(0)}=4\pi E_{X}a_{X}^{2},\label{eq:gXXBorn}
\end{equation}
which is about two times the exciton-exciton interaction strength
in Eq. (\ref{eq:gXX2D}) when a Coulomb interaction is considered.
To go beyond the Born approximation, we consider the GPF calculation
at a small light-matter coupling $\Omega=0.2E_{X}$ and a large photon
detuning $\delta=8E_{X}$, so the photon field is essentially not
populated and the system could be a perfect weakly interacting 2D
BEC of excitons in the dilute limit.

\begin{figure}[t]
\begin{centering}
\includegraphics[width=0.45\textwidth]{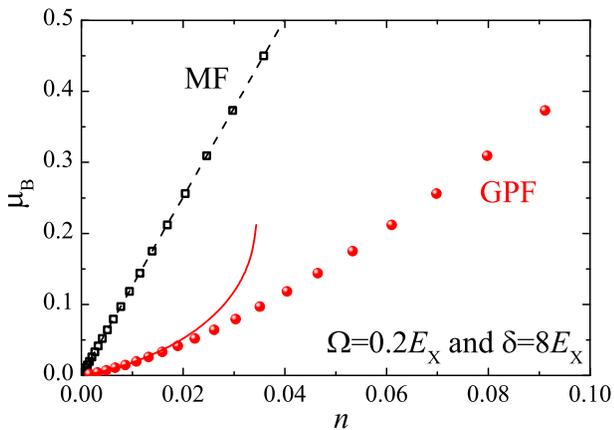} 
\par\end{centering}
\centering{}\caption{\label{fig1_ExcitonLimit} Bosonic chemical potential (in units of
$E_{X}$) as a function of the number density (in units of $a_{X}^{-2}$)
at strong light-matter coupling $\Omega=0.2E_{X}$ and at a large
photon detuning $\delta=8E_{X}$. The black empty squares and red
solid circles show the results obtained by mean-field and Gaussian
pair fluctuation theories, respectively. The black dashed line and
blue solid line are the predictions of the mean-field theory, Eq.
(\ref{eq:excitonMF}), and Bogoliubov theory for excitons, Eq. (\ref{eq:excitonGPF}),
in the dilute density limit, i.e., $y=(4\pi)x$ and $y=2\pi x/\left[-2\ln\kappa-(\ln2+1)-\ln x\right]\simeq2\pi x/(-0.5335-\ln x)$,
where $x\equiv na_{X}^{2}$ and $y\equiv\mu_{B}/E_{X}$. At the density
$n>0.02a_{X}^{-2}$, the GPF results cannot be explained by the Bogoliubov
theory. This is anticipated, since the gas parameter $na_{X}^{2}>0.02$
is already too large and the system is no longer in the weakly interacting
regime. }
\end{figure}

In Fig. \ref{fig1_ExcitonLimit}, we show the density equation of
state for small total density $n=n_{\textrm{tot}}$ or small chemical
potential $\mu_{B}=\mu-E_{LP}$, where $E_{LP}=-E_{X}$ is the energy
of the zero-momentum lower-polariton in the dilute limit in the absence
of the photon field. We find that the mean-field (empty squares) and
GPF results (solid circles) are indeed accurately described by Eq.
(\ref{eq:excitonMF}) and Eq. (\ref{eq:excitonGPF}), respectively.
We emphasize that, within the GPF theory, the chemical potential dependent
exciton-exciton interaction strength is given by, 
\begin{equation}
g_{XX}\left(\mu_{B}\right)=\frac{2\pi E_{X}a_{X}^{2}}{-2\ln\kappa-(\ln2+1)-\ln(\mu_{B}/E_{X})}.\label{eq:gXXBogoTheory}
\end{equation}
It vanishes logarithmically in the zero-density limit, i.e., $g_{XX}(\mu_{B}\rightarrow0)=0$.

\begin{figure}[t]
\begin{centering}
\includegraphics[width=0.45\textwidth]{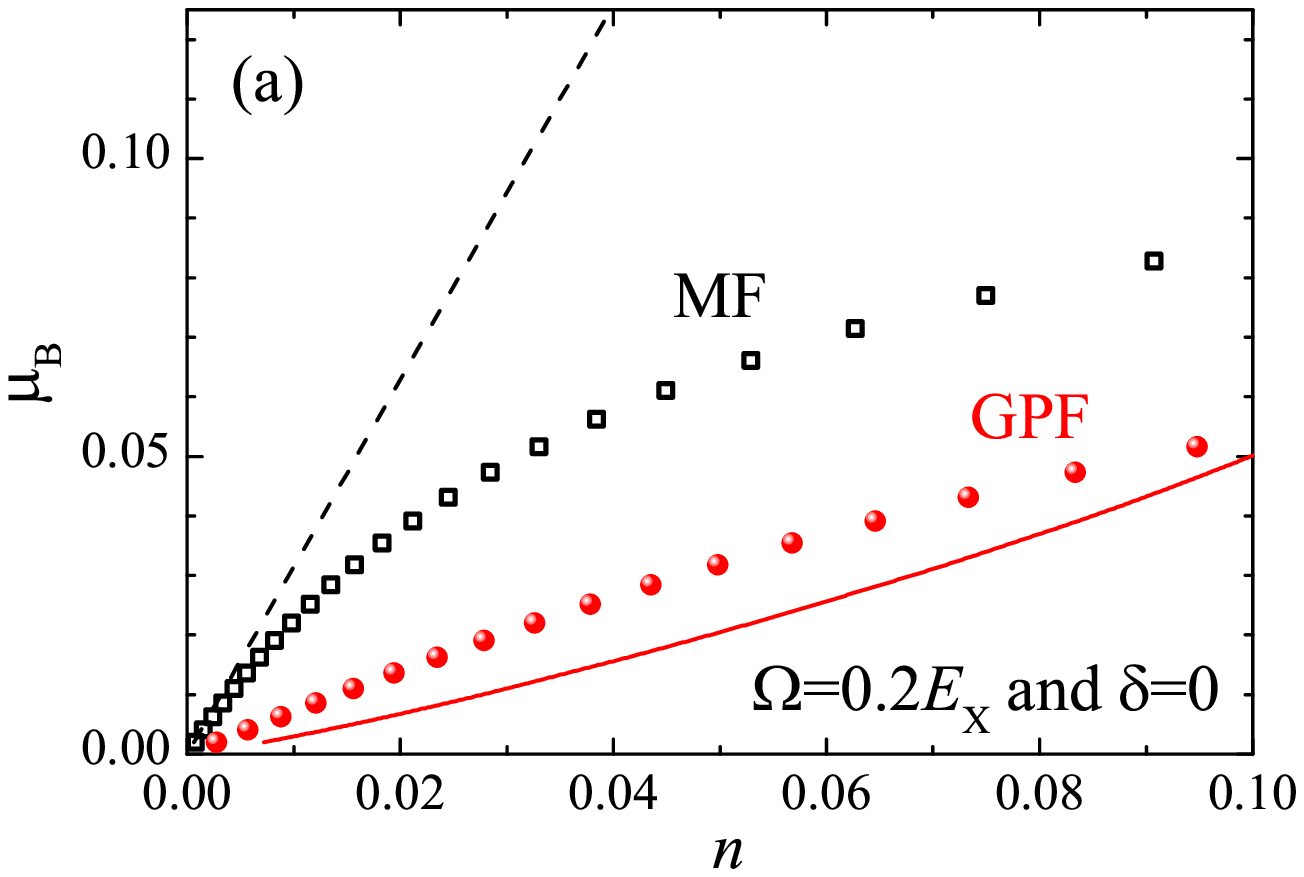} 
\par\end{centering}
\begin{centering}
\includegraphics[width=0.45\textwidth]{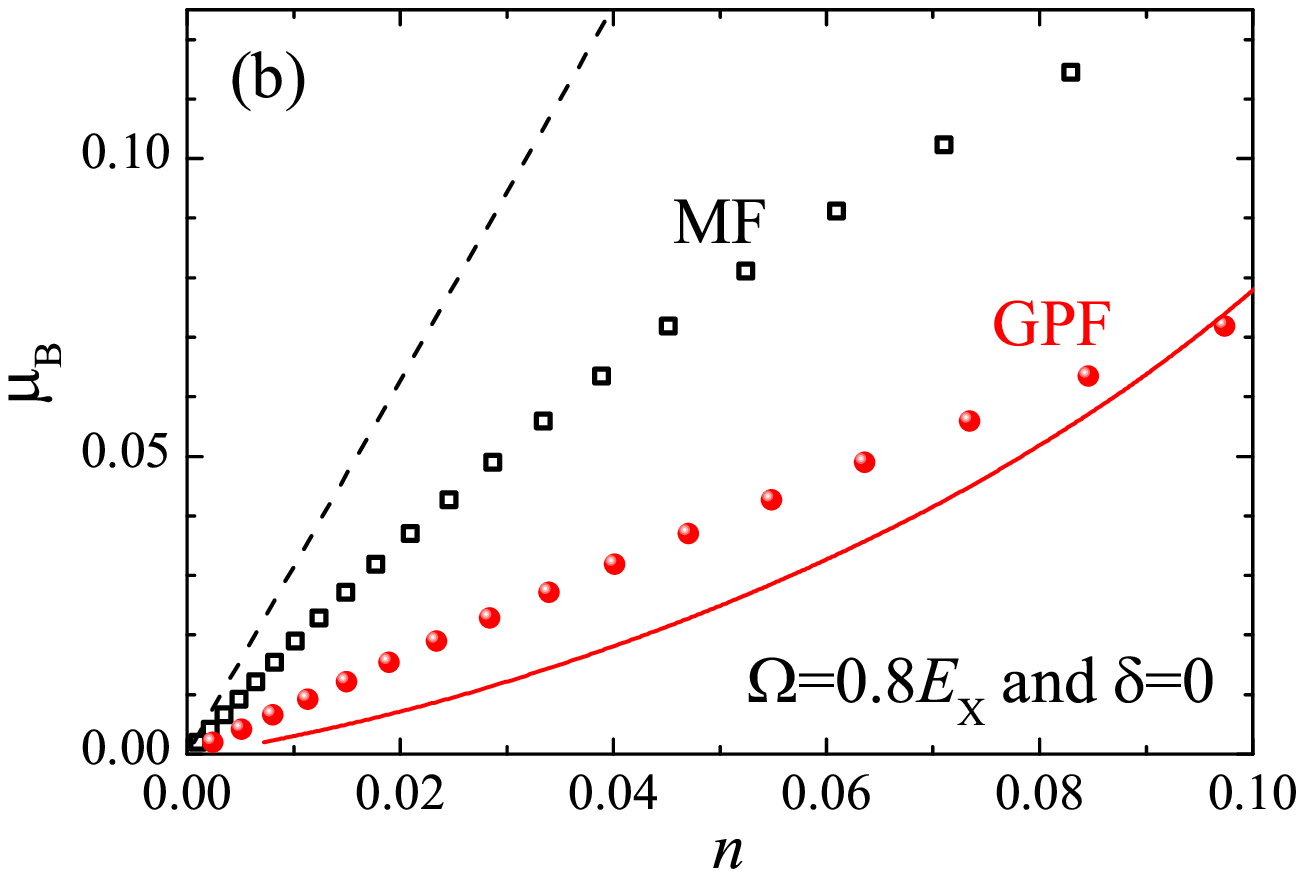} 
\par\end{centering}
\centering{}\caption{\label{fig2_PolaritonLimit} Bosonic chemical potential (in units
of $E_{X}$) as a function of the number density (in units of $a_{X}^{-2}$)
with zero photon detuning $\delta=0$, at strong light-matter coupling
$\Omega=0.2E_{X}$ (a) and at very strong light-matter coupling $\Omega=0.8E_{X}$
(b). The black empty squares and red solid circles show the results
obtained by mean-field and Gaussian pair fluctuation theories, respectively.
The black dashed line is the result from the Born approximation, $g_{PP}^{(0)}=X_{LP}^{4}(4\pi E_{X}a_{X}^{2})$.
The red solid line is based on the anticipation of a weakly interacting
2D Bose gas of exciton-polaritons, i.e., $g_{PP}=X_{LP}^{4}g_{XX}$,
where $g_{XX}$ is give by Eq. (\ref{eq:gXXBogoTheory}). Here, $X_{LP}^{2}=1/2$
at zero detuning according to the exciton-polariton model Eq. (\ref{eq:XLP2}). }
\end{figure}

\section{2D exciton-polariton condensate}

What happens if the photon field is significantly occupied? In Fig.
\ref{fig2_PolaritonLimit}, we show the mean-field and GPF density
equations of state at zero photon detuning $\delta=0$ and at two
light-matter couplings $\Omega=0.2E_{X}$ (a) and $\Omega=0.8E_{X}$
(b). For comparison, we show also the corresponding equations of state
predicted by the exciton-polariton model (see, i.e., Eq. (\ref{eq:gPP0EP}))
using black dashed line and red solid line, respectively. There are
two interesting observations. First, the mean-field result apparently
deviates from the anticipated behavior $g_{PP}^{(0)}=X_{LP}^{4}g_{XX}^{(0)}$
\cite{Tassone1999}, indicating the breakdown of the exciton-polariton
model. This deviation becomes larger when we increase the light-matter
coupling. On the other hand, the GPF result clearly shows a linear
dependence of the density on the chemical potential, suggesting the
existence of a \emph{constant} polariton-polariton interaction strength. 

\subsection{Born approximation (mean-field)}

Let us first analyze the mean-field results. From the mean-field thermodynamic
potential Eq. (\ref{eq:OmegaMF}), we may derive the gap equation,
\begin{equation}
\sqrt{\mu^{2}+4\Delta^{2}}-\mu=2\varepsilon_{0}\exp\left(\frac{4\pi\hbar^{2}}{Mu_{\textrm{eff}}}\right),
\end{equation}
and the number equation, 
\begin{equation}
n_{\textrm{MF}}=\left(\frac{g}{\tilde{\delta}-\mu}\right)^{2}\frac{\Delta^{2}}{u_{\textrm{eff}}^{2}}+\frac{M}{8\pi\hbar^{2}}\left(\sqrt{\mu^{2}+4\Delta^{2}}+\mu\right).
\end{equation}
In the dilute BEC limit, both the bosonic chemical potential $\mu_{B}=\mu-E_{LP}$
and the order parameter $\Delta$ are small controllable parameters,
compared with the low-polariton energy $E_{LP}\sim-E_{X}$. To the
leading order, we thus have 
\begin{equation}
u_{\textrm{eff}}\rightarrow u+\frac{g^{2}}{E_{LP}-\tilde{\delta}}\equiv u_{\textrm{LP}}.
\end{equation}
Taylor-expanding the gap equation, we find that, 
\begin{equation}
-E_{LP}-\mu_{B}-\frac{\Delta^{2}}{E_{LP}}=\varepsilon_{0}\exp\left(\frac{4\pi\hbar^{2}}{Mu_{\textrm{LP}}}\right)\left[1+\frac{\mu_{B}}{\mathcal{A}}\right],
\end{equation}
where 
\begin{equation}
\mathcal{A}^{-1}\equiv\frac{4\pi\hbar^{2}}{Mu_{\textrm{LP}}^{2}}\frac{g^{2}}{\left(\tilde{\delta}-E_{LP}\right)^{2}}.
\end{equation}
The leading term of the above gap equation is simply the expression
for the zero-momentum lower-polariton energy \cite{Hu2020}, i.e.,
$E_{LP}=-\varepsilon_{0}e^{4\pi\hbar^{2}/(Mu_{\textrm{LP}})}$. Using
this to eliminate the cut-off energy scale $\varepsilon_{0}$, we
obtain 
\begin{eqnarray}
-\frac{\Delta^{2}}{E_{LP}} & = & \left[1-\frac{E_{LP}}{\mathcal{A}}\right]\mu_{B}.\label{eq:gapBornApp}
\end{eqnarray}
Next, to the leading order the number equation can be casted into
the form, 
\begin{equation}
n_{\textrm{MF}}=\frac{M}{4\pi\hbar^{2}}\left[1-\frac{E_{LP}}{\mathcal{A}}\right]\left(-\frac{\Delta^{2}}{E_{LP}}\right).\label{eq:numBornApp}
\end{equation}
By using the fact that $n=n_{\textrm{tot}}=n_{\textrm{MF}}$ within
mean-field and by combining these two equations to remove the pairing
gap $-\Delta^{2}/E_{LP}$, we find that, 
\begin{equation}
\mu_{B}=\left[1-\frac{E_{LP}}{\mathcal{A}}\right]^{-2}\left(4\pi E_{X}a_{X}^{2}\right)n.
\end{equation}

It is readily seen that, the polariton-polariton interaction strength
within the mean-field (Born approximation) is given by, 
\begin{equation}
g_{PP}^{(0)}=\xi_{LP}^{4}\left(4\pi E_{X}a_{X}^{2}\right),\label{eq:gPP0ToyModel}
\end{equation}
where we have defined, 
\begin{equation}
\xi_{LP}^{2}\equiv\left[1-\frac{E_{LP}}{\mathcal{A}}\right]^{-1}=\left[1-\frac{4\pi\hbar^{2}}{Mu_{\textrm{LP}}^{2}}\frac{g^{2}E_{LP}}{\left(\tilde{\delta}-E_{LP}\right)^{2}}\right]^{-1}.\label{eq:xiLP2}
\end{equation}
By recalling that $4\pi E_{X}a_{X}^{2}=g_{XX}^{(0)}$ is the exciton-exciton
interaction strength for our toy model, we may interpret $\xi_{LP}^{2}$
as a \emph{generalized} exciton Hopfield coefficient. This interpretation
can be easily examined for a small light-matter coupling, at which
the exciton-polariton model is applicable. For a small Rabi coupling
$\Omega\ll E_{X}$, we may approximate $u_{\textrm{LP}}\simeq u$
and use the expression for the zero-momentum lower-polariton energy,
\begin{equation}
E_{LP}=-E_{X}+\frac{\delta}{2}-\sqrt{\frac{\delta^{2}}{4}+\Omega^{2}}.\label{eq:ELP}
\end{equation}
By further taking $g^{2}/u^{2}=M\Omega^{2}/(4\pi\hbar^{2}E_{X})$
and recalling that $\delta=\tilde{\delta}+E_{X}$, we find that, 
\begin{equation}
\xi_{LP}^{2}\simeq\left[1+\frac{\Omega^{2}}{\left(\tilde{\delta}-E_{LP}\right)^{2}}\right]^{-1}=X_{LP}^{2}.
\end{equation}
Thus, in the case of a small light-matter coupling, $\xi_{LP}^{2}$
reduces to $X_{LP}^{2}$, as we anticipate. An alternative explanation
for the \emph{generalized} exciton Hopfield coefficient $\xi_{LP}^{2}$
is given in Appendix A, where we consider the electron-hole vertex
function or the polariton Green function.

\begin{figure}[t]
\begin{centering}
\includegraphics[width=0.45\textwidth]{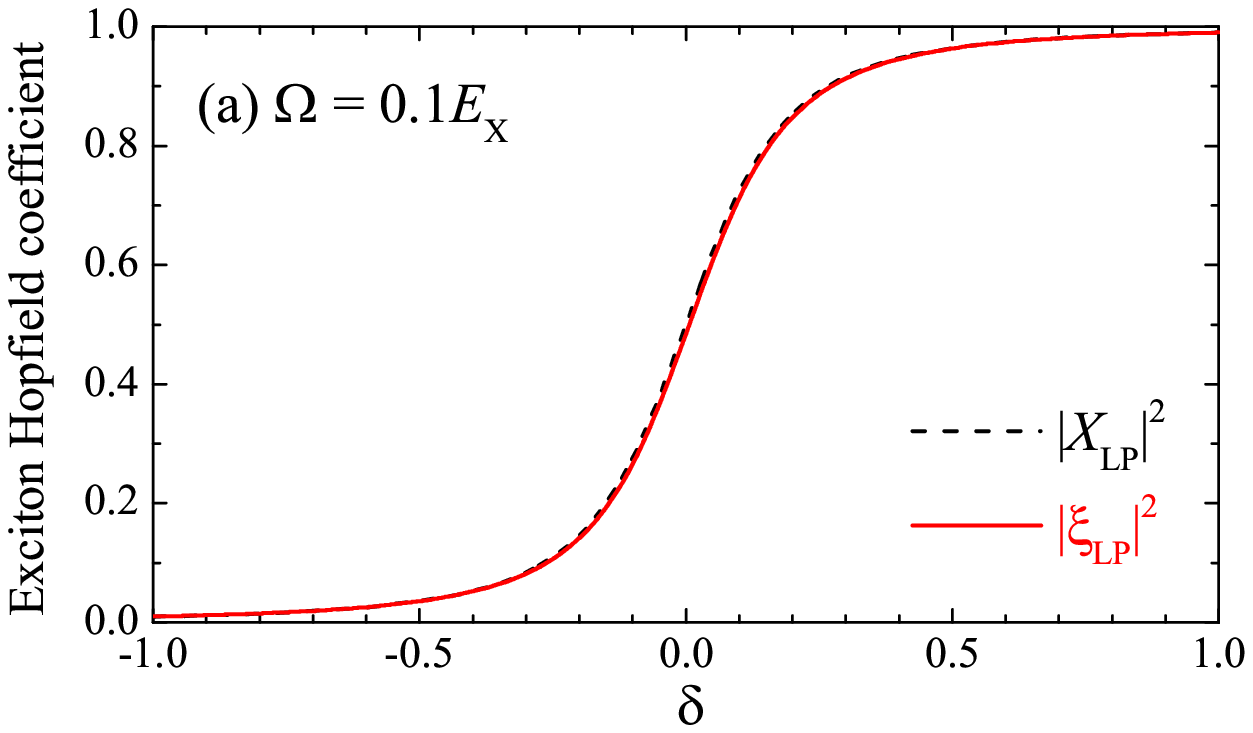} 
\par\end{centering}
\begin{centering}
\includegraphics[width=0.45\textwidth]{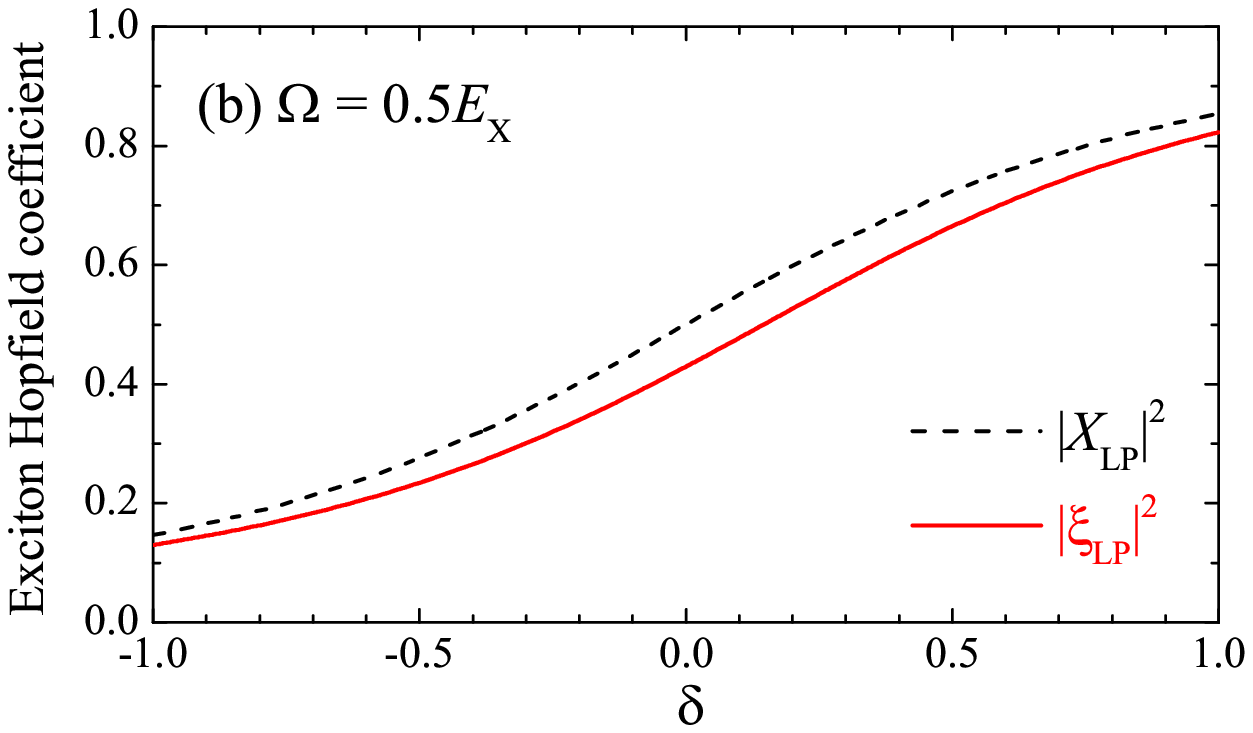} 
\par\end{centering}
\begin{centering}
\includegraphics[width=0.45\textwidth]{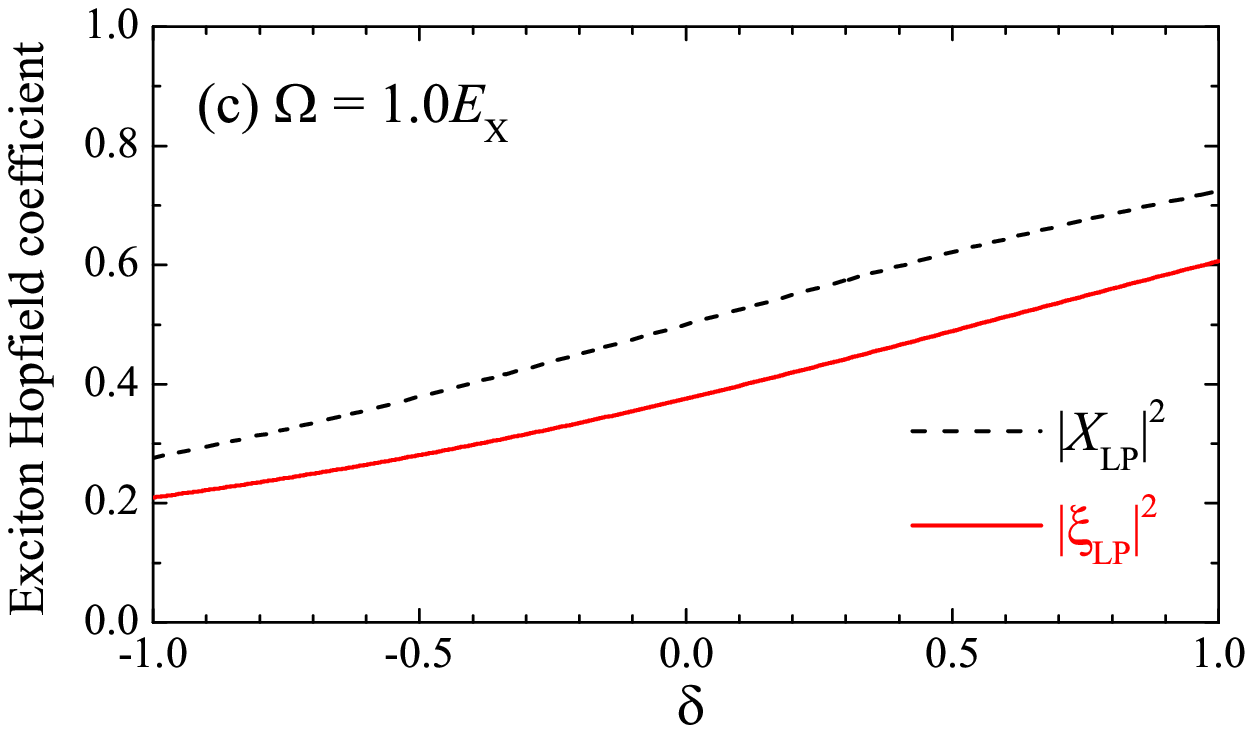} 
\par\end{centering}
\centering{}\caption{\label{fig3_XLP2} Exciton Hopfield coefficient $X_{LP}^{2}$ (black
dashed line) and the generalized exciton Hopfield coefficient $\xi_{LP}^{2}$
(red solid line) as a function of the photon detuning $\delta$ at
three light-matter couplings: $\Omega=0.1E_{X}$ (a), $0.5E_{X}$
(b), and $1.0E_{X}$ (c).}
\end{figure}

In Fig. \ref{fig3_XLP2}, we report the Hopfield coefficients $X_{LP}^{2}$
(black dashed line) and $\xi_{LP}^{2}$ (red solid line) as a function
of the photon detuning at three light-matter couplings: $\Omega=0.1E_{X}$
(a), $\Omega=0.5E_{X}$ (b), and $\Omega=1.0E_{X}$ (c). At small
coupling $\Omega\ll E_{X}$ as shown in (a), $\xi_{LP}^{2}$ is essentially
the same as the $X_{LP}^{2}$, as we have already confirmed analytically.
However, as the light-matter coupling increases, $\xi_{LP}^{2}$ becomes
increasingly smaller than $X_{LP}^{2}$ and the relative reduction
can be about a few $10\%$ when the light-matter coupling is comparable
to the exciton binding energy $\Omega\sim E_{X}$.

The difference between $\xi_{LP}^{2}$ and $X_{LP}^{2}$ at nonzero
light-matter coupling is expected. For the Coulomb interaction $V_{C}(r)\propto-1/r$,
it was understood in most previous works as the oscillator strength
saturation effect and its explicit form at the order of $\Omega/E_{X}$
was derived analytically \cite{Brichkin2011,Tassone1999}. The saturation
correction \emph{enhances} the polariton-polariton interaction strength.
This difference was also numerically investigated by Levinsen and
coworkers most recently \cite{Levinsen2019}. In addition to the known
saturation correction, a more dramatic effect of light-matter coupling
was revealed. At large light-matter coupling, the photon-mediated
attraction becomes dominant between electrons and holes \cite{Citrin2003}.
As a result, the size of excitons in the low-polariton branch shrinks
considerably and the exchange processes (for electrons or holes between
two different polaritons, which is responsible for polariton-polariton
repulsion) becomes less efficient \cite{Levinsen2019}. For our toy
model with a contact interaction between electrons and holes, the
reduction in the exchange processes seems to overwhelm the enhancement
due to the saturation in the oscillator strength, leading to an overall
smaller $\xi_{LP}^{2}$ in comparison with $X_{LP}^{2}$.

\subsection{Beyond the Born approximation (GPF)}

Here, we turn to consider the beyond-Born-approximation effect using
the GPF theory. Naïvely, we argue that the polariton system consists
of different types of carriers \cite{Hu2020}, as characterized by
$n_{\textrm{MF}}$ and $n_{\textrm{GPF}}$, which are contributed
from the mean-field saddle point and from pair fluctuations around
the saddle point, respectively. In the case of completely suppressed
fermionic degree of freedom, i.e., $n_{\textrm{MF}}\ll n_{\textrm{GPF}}$,
the system could be viewed as a weakly-interacting Bose gas of exciton-polaritons
and the density equation of state then follows the Bogoliubov theory,
as we have already discussed in Sec. III. This picture is not true
for the general case when the photon field starts to get occupied.
In general, as shown in Appendix B, we find that both $n_{\textrm{MF}}$
and $n_{\textrm{GPF}}$ become significant and towards the zero-density
limit, their ratio $n_{\textrm{MF}}/n_{\textrm{GPF}}$ saturates to
a constant. At large light-matter coupling and near zero photon detuning,
therefore, we may define a quantity, 
\begin{equation}
\mathcal{F}_{BB}=\lim_{\mu_{B}\rightarrow0}\left(\frac{n_{\textrm{MF}}}{n_{\textrm{tot}}}\right),
\end{equation}
which itself is functions of the light-matter coupling $\Omega$ and
of the photon detuning $\delta$. Now, using Eq. (\ref{eq:gapBornApp})
and Eq. (\ref{eq:numBornApp}) for $n_{\textrm{MF}}$, in the zero-density
limit we find, 
\begin{equation}
\mu_{B}=4\pi E_{X}a_{X}^{2}\xi_{LP}^{4}n_{\textrm{MF}}=\xi_{LP}^{4}\mathcal{F}_{BB}\left(4\pi E_{X}a_{X}^{2}\right)n,
\end{equation}
which implies a polariton-polariton interaction strength, 
\begin{equation}
g_{PP}=\left(\xi_{LP}^{4}\mathcal{F}_{BB}\right)\left(4\pi E_{X}a_{X}^{2}\right).\label{eq:gPPToyModel}
\end{equation}
In other words, within GPF the polariton-polariton interaction strength
is reduced by a factor of $\mathcal{F}_{BB}^{-1}$, compared with
the Born approximation result $g_{PP}^{(0)}=\xi_{LP}^{4}(4\pi E_{X}a_{X}^{2})$.
The linear dependence of the GPF result, as shown in Fig. \ref{fig2_PolaritonLimit}(b),
means that the mean-field contribution (i.e., the fermionic degree
of freedom and condensed photons) is significant. Otherwise, the reduction
factor $\mathcal{F}_{BB}$ will go to zero and the polariton-polariton
interaction strength $g_{PP}$ becomes zero. The polariton system
then crosses smoothly over to a weakly interacting 2D Bose gas of
exciton-polaritons, as we discuss in Sec. III.

\begin{figure}[t]
\begin{centering}
\includegraphics[width=0.45\textwidth]{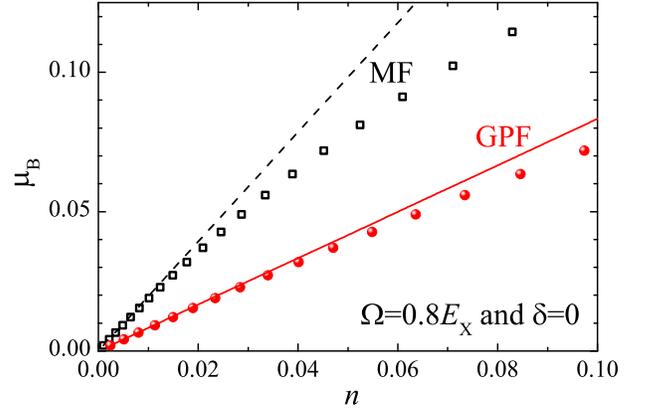} 
\par\end{centering}
\centering{}\caption{\label{fig4_gPPfitting} Bosonic chemical potential (in units of $E_{X}$)
as a function of the number density (in units of $a_{X}^{-2}$) with
zero photon detuning $\delta=0$ at very strong light-matter coupling
$\Omega=0.8E_{X}$. As the same as shown in Fig. \ref{fig2_PolaritonLimit}(b),
the black empty squares and red solid circles show the results obtained
by mean-field and Gaussian pair fluctuation theories, respectively.
But, now the black dashed line is the result from the Born approximation,
$g_{PP}^{(0)}=\xi_{LP}^{4}(4\pi E_{X}a_{X}^{2})$, with the generalized
exciton Hopfield coefficient $\xi_{LP}^{2}$. The red solid line shows
the result $g_{PP}=\xi_{LP}^{4}\mathcal{F}_{BB}(4\pi E_{X}a_{X}^{2})$,
which takes into account the reduction beyond the Born approximation.}
\end{figure}

\subsection{Comparison to the numerical results}

We can now understand the two observations made at the beginning of
this section, by using the main result of this work, 
\begin{equation}
\frac{g_{PP}}{g_{XX}^{(0)}}=\xi_{LP}^{4}\mathcal{F}_{BB},\label{eq:gPPMainResult}
\end{equation}
where $\xi_{LP}^{4}$ is responsible for the large light-matter coupling
and $\mathcal{F}_{BB}$ accounts for the beyond-Born-approximation
effect. In Fig. \ref{fig4_gPPfitting}, we replot Fig. \ref{fig2_PolaritonLimit}(b)
and add the anticipated behavior Eq. (\ref{eq:gPP0ToyModel}) for
the mean-field result (black dashed line) and Eq. (\ref{eq:gPPToyModel})
for the GPF result (red solid line). It is clear that in the low-density
limit, our analytic equations provide a satisfactory explanation to
the numerical results, obtained using either mean-field or GPF theories.

\begin{figure}[t]
\begin{centering}
\includegraphics[width=0.45\textwidth]{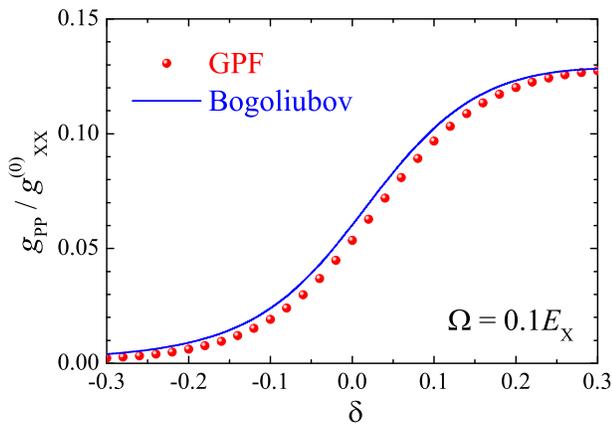} 
\par\end{centering}
\centering{}\caption{\label{fig5_gPPrabi01Bog} The ratio $g_{PP}/g_{XX}^{(0)}$ as a function
of the photon detuning $\delta$ (in units of $E_{X}$) at a light-matter
coupling $\Omega=0.1E_{X}$. The GPF result (red circles) is compared
with an analytic prediction from the exciton-polariton model within
the Bogoliubov theory (blue line), Eq. (\ref{eq:gPPBog}).}
\end{figure}

\subsection{Comparison to the analytic Bogoliubov result at small light-matter
coupling}

At small light-matter coupling, where the exciton-polariton model
is applicable, the polariton-polariton interaction strength can be
analytically obtained by using the Bogoliubov theory \cite{Hu2020arXiv}
or the scattering theory \cite{Bleu2020arXiv}. Taking the equal mass
for electrons and holes and the known exciton-exciton $s$-wave scattering
length $a_{s}=2\kappa e^{-\gamma}a_{X}$ (where $\kappa\simeq0.56$
as discussed in Sec. III) for a contact electron-hole attraction,
it takes the form \cite{Hu2020arXiv,Bleu2020arXiv}, 
\begin{equation}
\frac{g_{PP}}{g_{XX}^{(0)}}=\frac{X_{LP}^{4}}{2\ln\left[E_{X}/\left|E_{LP}^{(0)}\right|\right]-4\ln\left(2\kappa\right)},\label{eq:gPPBog}
\end{equation}
where $E_{LP}^{(0)}\equiv E_{LP}-(-E_{X})=\delta/2-\sqrt{\delta^{2}/4+\Omega^{2}}<0$
is the energy of zero-momentum lower-polariton, measured with respect
to the exciton energy $-E_{X}$. At small light-matter coupling, we
have $\xi_{LP}^{2}=X_{LP}^{2}$. Therefore, by comparing Eq. (\ref{eq:gPPMainResult})
and Eq. (\ref{eq:gPPBog}), we obtain that for $\Omega\ll E_{X}$,
\begin{equation}
\mathcal{F}_{BB}=\frac{1}{2\ln\left[E_{X}/\left|E_{LP}^{(0)}\right|\right]-4\ln\left(2\kappa\right)}.
\end{equation}
In Fig. \ref{fig5_gPPrabi01Bog}, we compare the numerical GPF result
and the analytic Bogoliubov prediction for the polariton-polariton
interaction strength (measured in units of $g_{XX}^{(0)}$) as a function
of the photon detuning at $\Omega=0.1E_{X}$. A good agreement is
found. Although two different theories with entirely different model
Hamiltonians (i.e., fermionic vs. bosonic) are used, both of them
reliably describe the exciton-polariton physics at small light-matter
coupling.

It is interesting to note that, Eq. (\ref{eq:gPPBog}) clearly shows
a pole at the lower-polariton energy $E_{LP}^{(0)}=-E_{X}/(4\kappa^{2})\simeq-0.8E_{X}$
or $\delta\simeq-0.8E_{X}$ under the condition $\Omega\ll E_{X}$.
This \emph{weak} logarithmic divergence is neutralized by the rapidly
decreasing excitonic Hopfield coefficient $X_{LP}^{4}\simeq(\Omega/\delta)^{4}\sim2.4\times10^{-4}$,
if we take $\Omega=0.1E_{X}$. As a result, the polariton-polariton
interaction strength $g_{PP}$ is always much smaller than the exciton-exciton
interaction strength $g_{XX}^{(0)}$ obtained within the Born approximation.
This situation, however, can dramatically change if the ratio $\kappa$
is allowed to tune experimentally (hopefully in transition-metal-dichalcogenide
monolayers \cite{Hu2020arXiv}). An enlarged ratio $\kappa$ shifts
the logarithmic pole in Eq. (\ref{eq:gPPBog}) to the zero photon
detuning $\delta\sim0$ and consequently the polariton-polariton interaction
strength $g_{PP}$ could be greatly enhanced. For more detailed discussions,
we refer to Ref. \cite{Hu2020arXiv}.

\section{Comparison to the experiment}

Although our main result Eq. (\ref{eq:gPPMainResult}) is obtained
by using a toy model Hamiltonian with a contact interaction for electrons
and holes, it would be interesting to see its relevance to the experimental
measurements, where a Coulomb-like interaction, i.e., Eq. (\ref{eq:VC}),
should be considered. To this aim, let us make a \emph{bold} assumption
that, Eq. (\ref{eq:gPPMainResult}) depends very weakly on the underlying
interaction between electrons and holes.

How can we assume that the beyond-Born-approximation effect should
lead to the \emph{same} reduction factor in the polariton-polariton
interaction strength, for both contact interaction and Coulomb interaction?
This is certainly difficult to justify. But, we may consider the exciton-exciton
interaction strength in 3D, which seems to be the only example available
for checking at the moment. According to a recent fixed-node diffusion
Monte Carlo simulation with Coulomb interaction in 3D \cite{Golomedov2017},
the exciton-exciton scattering length is about $a_{s}=1.5a_{X}$.
Here, for a single exciton, its ground state energy $E=-\hbar^{2}/(Ma_{X}^{2})$.
The Born approximation result for the exciton-exciton scattering length
can be extracted from the expression, 
\begin{equation}
g_{XX}^{(0)}=\frac{26\pi}{3}E_{X}a_{X}^{3}\equiv\frac{4\pi\hbar^{2}}{(2M)}a_{s}^{(0)},
\end{equation}
We therefore find that, $a_{s}^{(0)}=(13/3)a_{X}$. Thus, the ratio
between the exact result and the Born approximation result for the
exciton-exciton scattering length is about, 
\begin{equation}
\left[\frac{a_{s}^{(0)}}{a_{s}}\right]_{\textrm{Coulomb}}=\frac{13/3}{1.5}\simeq2.89.
\end{equation}
On the other hand, if we consider a contact interaction, the exact
exciton-exciton scattering length in 3D is $0.6a_{F}$ \cite{Petrov2004}
and the Born approximation result is $2a_{F}$, where $a_{F}$ is
the fermion-fermion scattering length in 3D, and we find that, 
\begin{equation}
\left[\frac{a_{s}^{(0)}}{a_{s}}\right]_{\textrm{contact}}=\frac{2}{0.6}\simeq3.33.
\end{equation}
The two ratios are surprisingly close, despite the entirely different
interaction potential between electrons and holes. This observation
may suggest that the reduction in the exciton-exciton interaction
strength or polariton-polariton interaction strength due to the beyond-Born-approximation
effect could be universal, depending weakly on the underlying interaction
between electrons and holes. We may then have a good reason to apply
our toy model results with a contact interaction.

\begin{figure}[t]
\begin{centering}
\includegraphics[width=0.45\textwidth]{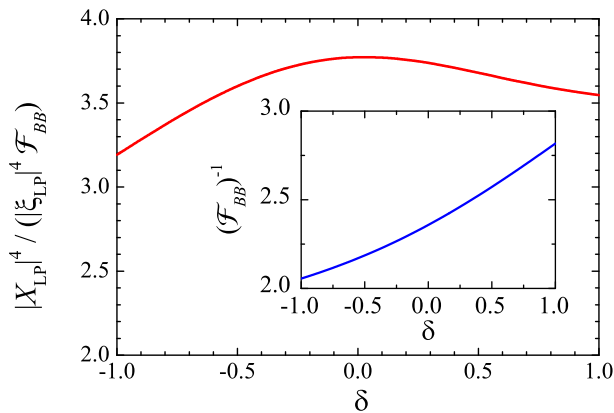} 
\par\end{centering}
\centering{}\caption{\label{fig6_BeyondBornRatio} The reduction factor in the polariton-polariton
interaction strength at a very strong light-matter coupling $\Omega=0.8E_{X}$
as a function of the photon detuning $\delta$, due to the combined
effects of the saturation in exciton oscillator strength and the beyond-Born-approximation
correction. The inset shows the inverse density fraction of fermionic
quasi-particles, $\mathcal{F}_{BB}^{-1}=n_{\textrm{tot}}/n_{\textrm{MF}}$,
as a function of the photon detuning $\delta$.}
\end{figure}

Therefore, it seems reasonable to consider a \emph{universal} ratio
defined by, 
\begin{equation}
\frac{g_{PP}}{X_{LP}^{4}g_{XX}^{(0)}}=\left(\frac{\xi_{LP}^{4}}{X_{LP}^{4}}\right)\mathcal{F}_{BB},
\end{equation}
which characterizes the two corrections: (i) the strong renormalization
to $X_{LP}^{4}$ due to a very strong light-matter coupling within
the Born approximation and (ii) the effect beyond the Born approximation.
In Fig. \ref{fig6_BeyondBornRatio}, we report the inverse of this
ratio as a function of the photon detuning at the light-matter coupling
$\Omega=0.8E_{X}$, at which the experimental data are taken. It is
about $3$ or $4$ upon changing the photon detuning. The most contribution
comes from the beyond-Born-approximation effect, as shown in the inset,
which gives about a factor of $2$ or $3$ reduction to the polariton-polariton
interaction strength.

\begin{figure}[t]
\begin{centering}
\includegraphics[width=0.45\textwidth]{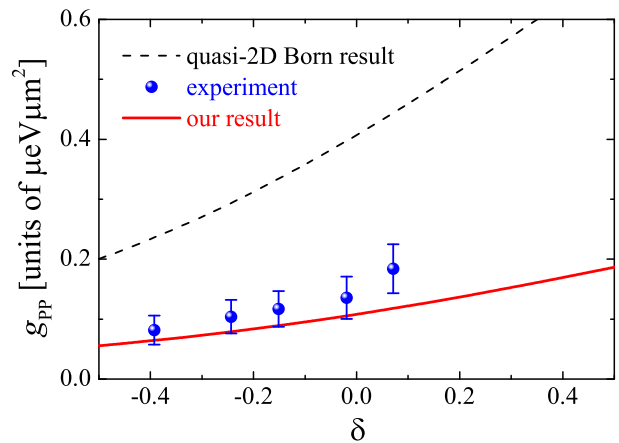} 
\par\end{centering}
\centering{}\caption{\label{fig7_TheoryVsExperiment} Theory versus experiment for the
polariton-polariton interaction strength at a very strong light-matter
coupling $\Omega\simeq0.8E_{X}$. Our beyond-Born-approximation prediction
(red solid line) is compared with the experimental data (blue circles
with error bars) that is taken from Fig. 5(a) in Ref. \cite{Estrecho2019}.
The black dashed line shows the result obtained with the Born approximation
in a quasi-2D configuration, i.e., Eq. (\ref{eq:gXXquasi2D}), together
with $X_{LP}^{2}$ calculated using Eq. (\ref{eq:XLP2}). Our beyond-Born-approximation
prediction is calculated by dividing the quasi-2D Born approximation
result by the reduction factor shown in Fig. \ref{fig6_BeyondBornRatio}.
It takes into account both the saturation effect in the exciton oscillator
strength and the correction beyond the Born approximation.}
\end{figure}

We can now multiply the ratio $(\xi_{LP}^{4}\mathcal{F}_{BB}/X_{LP}^{4})$
to the quasi-2D exciton-exciton interaction strength $g_{XX,\textrm{q2d}}^{(0)}$
in Eq. (\ref{eq:gXXquasi2D}), to obtain a reasonable estimate for
the polariton-polariton interaction strength. This is shown in Fig.
\ref{fig7_TheoryVsExperiment} using a red solid line, together with
the experimental data (blue dots with error bar) and the Born approximation
result $g_{PP}^{(0)}=X_{LP}^{4}g_{XX,\textrm{q2d}}^{(0)}$ that was
previously used as a theoretical upper bound (black dashed line).
By taking into account the factor of 3 or 4 reduction, our beyond-Born-approximation
theory seems to be in a reasonable agreement with the experimental
data.

\section{Conclusions and outlooks}

In conclusions, we have theoretically investigated the beyond-Born-approximation
effect for the polariton-polariton interaction based on a Gaussian
pair fluctuation theory \cite{Hu2020}, by using a toy model Hamiltonian
with a contact interaction for electrons and holes. This simplified
toy model enables us to understand the appearance of a constant polariton-polariton
interaction strength, which is usually assumed in previous studies
but is not theoretically guaranteed following the picture of a weakly
interacting two-dimensional Bose gas of exciton-polaritons. We have
shown that the effect beyond the Born approximation can lead to a
factor of 3 reduction in the polariton-polariton interaction strength.
As a by-product, the simplification also allows us to analytically
define a generalized exciton Hopfield coefficient, Eq. (\ref{eq:xiLP2}),
which takes into account the correction to the polariton-polariton
interactions at large light-matter coupling. We have made an attempt
to use our beyond-Born-approximation theory to understand the latest
experimental data of the polariton-polariton interaction strength
\cite{Estrecho2019}. A reasonable agreement has been found.

Future work will solve the exciton-exciton and polariton-polariton
interaction strengths under the Coulomb-like interaction Eq. (\ref{eq:VC}).
The results within the Born approximation should be easy to obtain.
We may simply generalize the work by Levinsen and his collaborators
\cite{Levinsen2019}, paying specific attention to the renormalization
of the light-matter coupling, as the exciton wave-functions are no
longer analytically available. Going beyond the Born approximation
will be very challenging. But, for the exciton-exciton interaction
strength, at least we may try solving the four-particle problem (two
electrons and two holes) in a numerically efficient way, using either
fixed-node Monte Carlo simulation as in three dimensions \cite{Golomedov2017}
or explicitly correlated Gaussian basis expansion approach \cite{Yin2019,Yin2020}. 
\begin{acknowledgments}
We thank Elena Ostrovskaya, Eliezer Estrecho, Maciej Pieczarka, Jesper
Levinsen, Meera Parish and Jia Wang for helpful discussions. This
research was supported by the Australian Research Council's (ARC)
Discovery Program, Grant No. DP170104008 (H.H.) and Grant No. DP180102018
(X.-J.L), and by the Army Research Office under Awards W911NF-17-1-0312
(H.D.). 
\end{acknowledgments}

\appendix

\section{Generalized exciton Hopfield coefficient}

We may clarify the physical meaning of the generalized exciton Hopfield
coefficient from the electron-hole pair vertex function in vacuum
$\Gamma_{\textrm{vac}}[\mathcal{Q}=(\mathbf{q},i\nu_{n})]$, which
takes the form, 
\begin{eqnarray}
\Gamma_{\textrm{vac}}\left(\mathcal{Q}\right) & = & \left[\frac{1}{u_{\textrm{eff}}\left(\mathcal{Q}\right)}+\Pi_{\textrm{vac}}\left(\mathcal{Q}\right)\right]^{-1},\\
 & \simeq & C\frac{\left|\xi_{LP}\left(\mathbf{q}\right)\right|^{2}}{i\nu_{n}-E_{LP}\left(\mathbf{q}\right)}.
\end{eqnarray}
The second equation in the above holds near the pole $i\nu_{n}\rightarrow E_{LP}(\mathbf{q})$,
with the constant $C$ and the generalized exciton Hopfield coefficient
$\xi_{LP}^{2}(\mathbf{q})$ to be determined. Let us focus on the
case $\mathbf{q}=0$ and recall that,
\begin{widetext}
\begin{equation}
\frac{1}{u_{\textrm{eff}}\left(\mathbf{q}=\mathbf{0},i\nu_{n}\right)}+\Pi_{\textrm{vac}}\left(\mathbf{q}=\mathbf{0},i\nu_{n}\right)=\left(u+\frac{g^{2}}{i\nu_{n}-\tilde{\delta}}\right)^{-1}-\frac{M}{4\pi\hbar^{2}}\ln\left(\frac{-i\nu_{n}}{\varepsilon_{0}}\right).
\end{equation}
By Taylor-expanding the right-hand-side of the above equation in terms
of the small quantity $x=i\nu_{n}-E_{LP}$, we find that, 
\begin{equation}
\frac{1}{u_{\textrm{eff}}\left(\mathbf{q}=\mathbf{0},i\nu_{n}\right)}+\Pi_{\textrm{vac}}\left(\mathbf{q}=\mathbf{0},i\nu_{n}\right)\simeq\left[\frac{g^{2}/\left(\tilde{\delta}-E_{LP}\right)^{2}}{\left(u+g^{2}/\left(\tilde{\delta}-E_{LP}\right)\right)^{2}}-\frac{M}{4\pi\hbar^{2}}\frac{1}{E_{LP}}\right]\left(i\nu_{n}-E_{LP}\right).
\end{equation}
Therefore, we obtain 
\begin{equation}
\Gamma_{\textrm{vac}}\left(\mathbf{q}=\mathbf{0},i\nu_{n}\right)\simeq\frac{4\pi\hbar^{2}}{M}\left(-E_{LP}\right)\frac{1}{i\nu_{n}-E_{LP}}\left[1-\frac{4\pi\hbar^{2}}{Mu_{\textrm{LP}}^{2}}\frac{g^{2}E_{LP}}{\left(\tilde{\delta}-E_{LP}\right)^{2}}\right]^{-1},
\end{equation}
implying 
\begin{eqnarray}
C & = & \frac{4\pi\hbar^{2}}{M}\left(-E_{LP}\right),\\
\xi_{LP}^{2}\left(\mathbf{q=0}\right) & = & \left[1-\frac{4\pi\hbar^{2}}{Mu_{\textrm{LP}}^{2}}\frac{g^{2}E_{LP}}{\left(\tilde{\delta}-E_{LP}\right)^{2}}\right]^{-1}.
\end{eqnarray}
\end{widetext}

\begin{figure}[t]
\centering{}\includegraphics[width=0.45\textwidth]{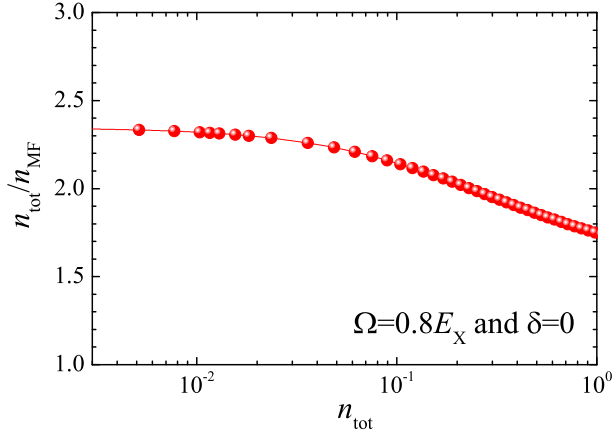}
\caption{\label{fig8_nttnmf} The ratio $n_{\textrm{tot}}/n_{\textrm{MF}}$
as a function of the total carrier density, at a very strong light-matter
coupling $\Omega=0.8E_{X}$ and at zero photon detuning $\delta=0$.}
\end{figure}

\section{Density dependence of the ratio $n_{\textrm{tot}}/n_{\textrm{MF}}$}

Here we discuss the ratio $n_{\textrm{tot}}/n_{\textrm{MF}}$ in the
low-density limit. As shown in Fig. \ref{fig8_nttnmf}, upon decreasing
total carrier density $n_{\textrm{tot}}$ (or effectively bosonic
chemical potential $\mu_{B}$), the ratio seems to saturate to a fixed
value, which depends on the light-matter coupling $\Omega$ and the
photon detuning $\delta$.


\begin{thebibliography}{10}
\bibitem{NoteShortRangeInteraction}We note that, although the underlying
interaction between electrons and holes is of the Coulomb type, the
effective interaction between composite bosons, including both excitons
and polaritons, could be described by a short-range Lennard-Jones
potential $W[(a_{*}/r)^{12}-(a_{*}/r)^{6}]$, with a strength $W$
and a length scale $a_{*}$ comparable to the excitonic Bohr radius
$a_{X}$. At low temperature, only the $s$-wave channel is important
and we then can use an $s$-wave scattering length $a_{s}$ to characterize
the effective interaction.
\bibitem{NoteBornApproximation}In scattering theory, the Born approximation
amounts to truncating the Lippmann-Schwinger equation to the first
order and taking the bare scattering potential to replace the two-body
$T$-matrix. In quantum many-body approach, the Born approximation
result for the inter-particle interaction strength can often be conveniently
obtained, by taking mean-field approximation to the many-body theory.
In this context, we take the Born approximation and mean-field approximation
as exchangeable terminology.
\bibitem{NotationNote}We note that, slightly different notations
are used in Ref. \cite{Hu2020}. There, the Bohr radius and binding
energy are denoted by $a_{B}$ and $E_{B}$, respectively. The photon
detuning $\delta$ is measured from the edge of the bandgap $E_{\textrm{g}}$
and should be identified as $\tilde{\delta}$ in the current manuscript.
\end{thebibliography}
\end{document}